\documentclass[usenatbib,useAMS]{mn2e}
\usepackage{amssymb, amsmath, epsf, graphicx}
\usepackage{deluxetable}
\usepackage[dvips]{color}

\def\jnl@style{\it}
\def\aaref@jnl#1{{\jnl@style#1}}

\def\aaref@jnl#1{{\jnl@style#1}}

\def\aj{\aaref@jnl{AJ}}                   
\def\araa{\aaref@jnl{ARA\&A}}             
\def\apj{\aaref@jnl{ApJ}}                 
\def\apjl{\aaref@jnl{ApJ}}                
\def\apjs{\aaref@jnl{ApJS}}               
\def\ao{\aaref@jnl{Appl.~Opt.}}           
\def\apss{\aaref@jnl{Ap\&SS}}             
\def\aap{\aaref@jnl{A\&A}}                
\def\aapr{\aaref@jnl{A\&A~Rev.}}          
\def\aaps{\aaref@jnl{A\&AS}}              
\def\azh{\aaref@jnl{AZh}}                 
\def\baas{\aaref@jnl{BAAS}}               
\def\jrasc{\aaref@jnl{JRASC}}             
\def\memras{\aaref@jnl{MmRAS}}            
\def\mnras{\aaref@jnl{MNRAS}}             
\def\pra{\aaref@jnl{Phys.~Rev.~A}}        
\def\prb{\aaref@jnl{Phys.~Rev.~B}}        
\def\prc{\aaref@jnl{Phys.~Rev.~C}}        
\def\prd{\aaref@jnl{Phys.~Rev.~D}}        
\def\pre{\aaref@jnl{Phys.~Rev.~E}}        
\def\prl{\aaref@jnl{Phys.~Rev.~Lett.}}    
\def\pasp{\aaref@jnl{PASP}}               
\def\pasj{\aaref@jnl{PASJ}}               
\def\qjras{\aaref@jnl{QJRAS}}             
\def\skytel{\aaref@jnl{S\&T}}             
\def\solphys{\aaref@jnl{Sol.~Phys.}}      
\def\sovast{\aaref@jnl{Soviet~Ast.}}      
\def\ssr{\aaref@jnl{Space~Sci.~Rev.}}     
\def\zap{\aaref@jnl{ZAp}}                 
\def\nat{\aaref@jnl{Nature}}              
\def\iaucirc{\aaref@jnl{IAU~Circ.}}       
\def\aplett{\aaref@jnl{Astrophys.~Lett.}} 
\def\apspr{\aaref@jnl{Astrophys.~Space~Phys.~Res.}}
\def\bain{\aaref@jnl{Bull.~Astron.~Inst.~Netherlands}} 
\def\fcp{\aaref@jnl{Fund.~Cosmic~Phys.}}  
\def\gca{\aaref@jnl{Geochim.~Cosmochim.~Acta}}   
\def\grl{\aaref@jnl{Geophys.~Res.~Lett.}} 
\def\jcp{\aaref@jnl{J.~Chem.~Phys.}}      
\def\jgr{\aaref@jnl{J.~Geophys.~Res.}}    
\def\jqsrt{\aaref@jnl{J.~Quant.~Spec.~Radiat.~Transf.}}
\def\memsai{\aaref@jnl{Mem.~Soc.~Astron.~Italiana}}
\def\nphysa{\aaref@jnl{Nucl.~Phys.~A}}   
\def\physrep{\aaref@jnl{Phys.~Rep.}}   
\def\physscr{\aaref@jnl{Phys.~Scr}}   
\def\planss{\aaref@jnl{Planet.~Space~Sci.}}   
\def\procspie{\aaref@jnl{Proc.~SPIE}}   

\def\eps@scaling{1.0}%
\newcommand\epsscale[1]{\gdef\eps@scaling{#1}}%
\newcommand\plotone[1]{%
 \centering
 \leavevmode
 \includegraphics[width={\eps@scaling\columnwidth}]{#1}%
}%

\newcommand{\lum}{ergs s\ensuremath{^{-1}}}
\newcommand{\lbol}{\ensuremath{L\mathrm{_{bol}}}}
\newcommand{\ledd}{\ensuremath{L\mathrm{_{Edd}}}}
\newcommand{\lratio}{\lbol/\ledd}
\newcommand{\lfive}{\ensuremath{L_{5100}}}
\newcommand{\msun}{\ensuremath{M_{\odot}}}

\newcommand{\mbh}{\ensuremath{M_\mathrm{BH}}}

\newcommand{\percubecm}{\ensuremath{\mathrm{cm}^{-3} }}
\newcommand{\indw}{\ensuremath{\alpha_{\lambda}}}

\newcommand{\chisq}{\ensuremath{\chi^2}}
\newcommand{\aox}{\ensuremath{\alpha_{\rm{ox}}}}

\newcommand{\ha}{H\ensuremath{\alpha}}
\newcommand{\hb}{H\ensuremath{\beta}}
\newcommand{\hc}{H\ensuremath{\gamma}}
\newcommand{\hd}{H\ensuremath{\delta}}
\newcommand{\he}{H\ensuremath{\epsilon}}

\newcommand{\bha}{H\ensuremath{\alpha ^{b}}}

\newcommand{\bhb}{H\ensuremath{\beta  ^{b}}}

\newcommand{\bhc}{H\ensuremath{\gamma ^{b}}}

\newcommand{\bhd}{H\ensuremath{\delta ^{b}}}

\newcommand{\nii}{[N\,II]}
\newcommand{\sii}{[S\,II]}
\newcommand{\oiii}{[O\,III]}
\newcommand{\feii}{Fe\,II}

\title[Broad-line Balmer Decrements]{Broad-line Balmer Decrements in Blue
Active Galactic Nuclei}

\author[X.~Dong et al.]{\parbox[t]{\textwidth}{
Xiaobo Dong$^{1,2}$,
Tinggui Wang$^{1,2}$,
Jianguo Wang$^{3,4}$,
Weimin Yuan$^{3}$,
Hongyan Zhou$^{1,2,5}$,
Haifeng Dai$^{1,2}$ and Kai Zhang$^{1,2}$}\\
\parbox[t]{\textwidth}
{$^1$Center for Astrophysics, University of Science and
Technology of China (USTC), Hefei, Anhui, 230026, China;~
xbdong, twang@ustc.edu.cn}\\
\parbox[t]{\textwidth}{
$^2$Joint Institute of Galaxies and Cosmology,
Shanghai Observatory and USTC}\\
\parbox[t]{\textwidth}{
$^3$National Astronomical Observatories/Yunnan Observatory,
Chinese Academy of Sciences, P.O. Box 110, Kunming, Yunnan 650011, China;~
jgwang, wmy@ynao.ac.cn}\\
\parbox[t]{\textwidth}{
$^4$Department of Physics, Yunnan University, 650031, Kunming, China}\\
\parbox[t]{\textwidth}{$^5$Max-Planck-Institut f\"ur extraterrestrische Physik,
Giessenbachstrasse 1, 85748 Garching, Germany}
}

\begin{document}
\maketitle \label{firstpage}

\begin{abstract}
We have investigated the broad-line Balmer decrements (\ha/\hb)
for a large, homogeneous sample of Seyfert 1 galaxies and QSOs
using spectroscopic data obtained in the Sloan Digital Sky Survey.
The sample, drawn from the Fourth Data Release,
comprises 446 low redshift ($z \lesssim 0.35$) active galactic nuclei (AGN)
that have blue optical continua as indicated by the spectral slopes
in order to minimize the effect of dust extinction.
We find that (i) the distribution of the intrinsic broad-line \ha/\hb\ ratio
can be well described by log-Gaussian, with a peak at \ha/\hb\ =3.06
and a standard deviation of about 0.03 dex only;
(ii) the Balmer decrement does not correlate
with AGN properties such as luminosity, accretion rate, and continuum slope, etc.;
(iii) on average, the Balmer decrements are found to be only slightly larger
in radio-loud sources (3.37) and
sources having double-peaked emission-line profiles (3.27)
compared to the rest of the sample.
We therefore suggest that the
broad-line \ha/\hb\ ratio can be used as a good indicator for
dust extinction in the AGN broad-line region;
this is especially true for radio-quiet AGN with regular emission-line profiles,
which constitute the vast majority of the AGN population.

\end{abstract}


\section{Introduction}

Hydrogen Balmer decrements are often used to determine the amount of dust extinction
attenuating the observed emission lines because the intrinsic decrements of Balmer
recombination lines are quite insensitive to the gas temperature and density
in low density, dilute radiation field conditions (Osterbrock 1989).
Particularly, the \ha/\hb\ ratio is most frequently used
for the strongness of the \ha\ and \hb\ lines and
their relatively large wavelength span,
which render the derived amount of extinction
less affected by measurement uncertainties.
For HII region photoionized by a hot star,
an intrinsic \ha/\hb\ value of 2.87 is found,
as predicted by Case B recombination
(at a typical electron density of $\lesssim10^4$ cm$^{-3}$
and temperature of $10^4$ K).
A value of 3.1 is generally adopted for the narrow line region (NLR)
of active galactic nuclei (AGN), where
\ha\ emission is slightly enhanced by collisional excitation
due to the presence of gas of higher densities
and the presence of partly ionized transition region resulting from
much harder ionizing continuum
(Gaskell \& Ferland 1984, Halpern \& Steiner 1983).
For the broad line region (BLR) of AGN,
however, where the density is so high
(typically $n_e \gtrsim 10^9$\percubecm) that collisional, optical-depth
and radiative-transfer effects become important
as predicted by BLR photoionization
modeling (e.g., Netzer 1975, Kwan \& Krolik 1981, Canfield \& Puetter 1981,
Collin-Souffrin et al. 1982, Rees et al. 1989, Dumont et al. 1998,
Korista \& Goad 2004, see Baldwin 1997 for a review),
the broad-line \ha/\hb\ ratio (\bha/\bhb) varies widely in
different BLR conditions.
Observationally, the \bha/\bhb\ ratios are larger (steeper)
than the Case B recombination value in most Seyfert 1s and QSOs
(e.g., Osterbrock 1977, Wu et al. 1980, Rafanelli 1985,
and Fig. 3 of Dong et al. 2005);
moreover, in broad-line radio galaxies and Seyfert 1.8/1.9 galaxies
\bha/\bhb\ can be as steep as 10 or higher
(Osterbrock et al. 1976, Osterbrock 1981, Crenshaw et al. 1988).
Thus, although it is unclear yet whether
the observed steep \bha/\bhb\ is due to the high-density effects
mentioned above, or just to wavelength-dependent extinction by dust
(e.g., Osterbrock 1984, Goodrich 1995),
it has been generally believed that the \bha/\bhb\ ratio
cannot be used as an indicator of reddening in BLR,
as is in AGN NLRs or HII regions.

However, we have noticed that the range of the \bha/\bhb\ ratios
of Seyfert 1/QSOs is fairly small,
typically from 2.5 to 5 (see references above),
probably suggestive of a small dispersion of the intrinsic \bha/\bhb\ ratio.
For instance, for 94 Seyfert 1s and QSOs having
$u^{\prime}-g^{\prime}\lesssim 0.6$
culled from the Sloan Digital Sky Survey (SDSS; York et al. 2000)
Early Data Release (EDR; stoughton et al. 2002),
the standard deviation of \ha/\hb\ is only 0.36 around a mean of 2.98
(Dong et al. 2005).
This fact motivated us to explore
systematically the intrinsic Balmer decrements in AGN,
their distribution and potential dependence on
other AGN properties (e.g. radio-loudness, luminosity,
accretion rate, line-profile parameters, continuum slope),
by taking advantage of the unprecedented spectroscopic data from the SDSS.
This study will be able to address the question as to
whether the \bha/\bhb\ ratio can be taken as an indicator of
the BLR extinction, at least in a {\em statistical} manner
and for specific sub-classes of AGN.
To this end, we have compiled a larger sample of blue AGN,
including both Seyfert 1s and QSOs,
from the SDSS Fourth Data Release (Adelman-McCarthy et al. 2006)
as to the shape of their continua.
We assume a cosmology with H$_{0}$=70 km~s$^{-1}$~Mpc$^{-1}$,
$\Omega_{M} $=0.3 and $\Omega_{\Lambda}$=0.7.

\section{Sample Construction and Data Analysis}

\subsection{Sample Definition}

Our aim is to select blue AGN that are free of dust extinction.
It has been noted that even in quasars with relatively blue colors, such as
Palomar-Green (PG) quasars ($U-B<-0.44$, Schmidt \& Green 1983),
there is noticeable internal dust extinction inside the AGN
(Rowan-Robinson 1995, Baskin \& Laor 2004).
A color criterion bluer than the average of QSOs is desirable.
The average optical--near-ultraviolet slope of QSO continuum
is found to be $\indw \thickapprox 1.5$
($f_{\lambda} = \lambda ^{-\indw}$;
Vanden Berk et al. 2001 and references listed in their Table 5).
We thus define blue AGN as those with a continuum slope
$\indw \gtrsim 1.5$, where
\indw\ is fitted in the rest-wavelength range of 4030--5600 \AA.
We consider thus selected AGN
to be least affected by dust extinction.
This point is to be further discussed in \S4.1,
as well as the representativeness of our sample.
In practice, we limit redshifts $z\lesssim 0.35$
so that the \ha\ line lies within the SDSS spectral coverage.
We select objects with the median spectral
signal-to-noise ratio (S/N) $\geq 10$ per pixel only
to ensure accurate measurement of the Balmer decrement.

\subsection{Overview of Data Processing}
Applying the above redshift and S/N cutoffs yielded a pool of
$\sim$4100 objects classified as AGN in the SDSS DR4 spectral data set,
from which our sample is to be culled.
The spectra are first corrected for Galactic extinction using the extinction
map of Schlegel et al.\ (1998) and
the reddening curve of Fitzpatrick (1999).
The spectra are transformed into the
rest frame using the redshift as
determined from the peak of the [O III]$\lambda5007$ emission line.
In order to measure accurately broad Balmer lines,
we have to subtract properly  the continuum,
the Fe\,II emission multiplets, and other emission lines nearby.
In common practice, the subtraction is performed step by step:
first to fit and subtract the
AGN continuum and then Fe\,II multiplets (or the opposite)
and finally to fit emission lines
(e.g., Boroson \& Green 1992, Marziani et al. 2003);
or first to fit and subtract
simultaneously the continuum and the Fe\,II emission multiplets
(so-called ``pseudo-continuum'') and then to fit other emission lines
(e.g., Dong et al. 2005, Greene \& Ho 2005, Zhou et al. 2006).
Unfortunately, for the optical spectra of most Seyfert 1s and QSOs,
fitting the continuum and the Fe\,II emission
is highly complicated by several facts as follows.
1) There are essentially no emission-line--free regions where
the continuum can be determined
(e.g., Vanden Berk et al. 2001).
2) The Fe\,II $\lambda\lambda4434-4684$ features,
generally prominent,
are often blended with broad lines of \hc, He\,II $\lambda4686$ and \hb.
3) Often the QSO continuum cannot be described by a single power-law
from \hd\ to \ha, which means that we have to determine
the local continuum for the \hb\ and \ha\ regions separately.
Limited by these complications,
the common step-by-step spectral subtraction procedure
cannot achieve Balmer decrement measurement
accurate enough for our purpose in this work,
which is rather sensitive to
the measurement uncertainties of the emission-line fluxes\footnote{
For instance, with a typical error of 10\%
for the emission-line fluxes as in Osterbrock 1977,
the error of the \ha/\hb\ ratios is 14\% (0.061 dex),
which is fairly large for our purpose (cf. \S3.1). }.
Here we adopt an alternative method to fit simultaneously the continuum,
the Fe\,II and the other emission lines,
giving emphasis on proper determination of the local pseudo-continua.
If there are still large residuals left in the \hb\ and/or \ha\ regions,
a refined fit of the emission-lines is performed to the
pseudo-continuum subtracted spectra.

Having completed the spectral fitting procedure,
we select blue AGN according to the above slope criterion.
Several objects having many bad pixels in the \hb\ or \ha\ regions are removed.
For eleven objects having duplicated spectra, we retain
the one tagged as `SciencePrimary' only according to
the SDSS Catalog Archive Server\footnote{http://cas.sdss.org/}.
Our final sample is composed of 446 Seyfert 1s and QSOs.
Our spectral fitting method is described in detail in \S2.3 and 2.4.

\subsection{Simultaneous Fit of Continuum and Emission Lines} 
As discussed above, we need to fit simultaneously
the respective local continua in the \hb\ and \ha\ regions
and the Fe\,II emission spectrum.
We follow the procedure described in below, which is implemented using
IDL routines:\footnote{
We use the MPFIT package for nonlinear fitting.
MPFIT is kindly provided by Craig B. Markwardt, available
at http://cow.physics.wisc.edu/\~{}craigm/idl/.}

1. We fit each SDSS spectrum in the rest-wavelength range
from 4030\AA\ to 7500\AA, assuming a broken power law
with a break wavelength of 5600\AA,
i.e., $a_{1} \lambda ^{-\alpha_{\lambda,1}}$
for the \hd--\hc--\hb\ region and
$a_{2} \lambda ^{-\alpha_{\lambda,2}}$ for the \ha\ region.
We limit the wavelength range
redward of 4030\AA\ to avoid the broad-line \he\ emission.

2. The optical Fe\,II emission is modeled as
$C(\lambda) = c_{b}C_{b}(\lambda) + c_{n}C_{n}(\lambda)$,
where $C_{b}(\lambda)$ represents the broad Fe\,II lines with
the relative intensities fixed at those of I\,ZW\,1 as given in
Table A.1 of V{\'e}ron-Cetty et al.\ (2004),
and with the same line profile as the broad \hb\ line.
The redshift of the broad  Fe\,II lines relative to \hb\
is fitted as a free parameter.
$C_{n}(\lambda)$ denotes the narrow permitted and forbidden
Fe\,II lines with their relative intensities fixed at those
listed in Table A.2 of V{\'e}ron-Cetty et al (2004);
which have the same redshifts and line profiles
as the narrow \hb\ line.
\footnote{Instead of the template spectrum of the broad Fe\,II emission of I\,ZW\,1
readily provided by V{\'e}ron-Cetty et al.\ (2004),
we use two sets of Fe\,II emission templates in analytical forms
constructed from their measurements,
one for the broad line system L1 and
the other for low-excitation narrow line system N3.
In addition, we also add into our templates Ti\,II, Ni\,II, Cr\,II lines
listed in their Table A.1 and A.2.}

3. Emission lines other than Iron lines identified from the composite
SDSS QSO spectrum (see Table 2 in Vanden Berk et al. 2001)
from \hd\ to [S\,II] $\lambda6731$\footnote{
Several emission-line regions in the 4030--7500\AA\ range
are simply masked out,
because either they are too weak to constrain in the fit
or they have little effect on the results.
These include He\,I $\lambda4471$,
[N\,I] $\lambda5200$, [Ca V] $\lambda5310$, [Cl III] $\lambda5538$,
He\,I $\lambda5876$, He\,I $\lambda7066$, [Ar III] $\lambda7136$
and [O\,II] $\lambda7320$.}
are modeled as follows.
Broad Hydrogen Balmer lines are assumed to have
the same redshifts and profiles, and each is modeled with 1--4 Gaussians.
The broad He\,II $\lambda4686$ line is modeled with one Gaussian.
[O\,III] $\lambda4363$ and the $\lambda\lambda4959,5007$ doublet
are assumed to have the same redshifts and profiles,
and each is modeled with 1--2 Gaussians.
Other narrow lines are modeled with one Gaussian.
Narrow Balmer lines, [N\,II] and [S\,II] doublets
are assumed to have the same redshift and profile.
The flux ratio of the [N\,II] doublet
$\lambda$6583/$\lambda$6548 is fixed to the theoretical value of 2.96;
the [O\,III] doublet is similarly constrained.

We note that in many cases the fit is not good in several wavelength regions,
such as the small region around the minimum between
\feii\ 37,38 complex and \hb,
the joint part of \hb\ and [O\,III] $\lambda$4959.
These are likely caused by the imperfection of the models,
for instance, over-estimation
of the continuum between the \feii\ 37,38 complex and \hb,
over-fit of \hb\ with a spurious very broad component,
over-fit of \oiii\ with a spurious blue-shifted wing.
These problems can be solved by assigning additionally larger weights to
these critical regions in the fitting, following the weighting methods
adopted by Tran et al. (1992) and Reichard et al. (2003).
The exact weights are determined by trial-and-error
as those which give the best fits---with the minimum $\chi^2$ calculated
using the \emph{original} weights (errors)---in the relevant regions.

We also note that the broad lines of He\,I $\lambda$4922 and $\lambda$5016
may contribute to the ``red shelf'' of \hb\ (e.g., V{\'e}ron et al. 2002).
To investigate this possibility, we compare the \hb\ profiles
with the \ha\ profiles in our sample.
Only 4 per cent of the sources are found to
have apparent redward excess of \hb, and
the strength is typically less than 5 per cent of \hb.
We thus consider the contribution of the potential He\,I lines
to be negligible.

The assumption that the broad Balmer lines have the same profiles
is useful to well constrain these lines in the fitting since
they are highly blended with other lines nearby.
This assumption is found to have little effect on the
determination of the local continua (see \S2.2),
although it is well known that broad Balmer lines
have slightly different profiles (e.g. Osterbrock \& Shuder 1982).
A similar situation arises in the comparison of the [O\,III]$\lambda$4363
and $\lambda$5007 line profiles, as the former being often broader.

We notice that in many objects the relative intensities of
various Fe\,II multiplets, such as $\frac{FeII\,48,49}{FeII\,37,38}$,
are more or less different from those of I\,ZW\,1,
resulting apparent residuals left in some Fe\,II multiplet regions.
For about two dozen such spectra we re-fit them
with different scaling factors for the
Fe\,II emission blueward and redward of \hb.
This yields better matched Fe\,II emission spectrum;
the local continua and the \hb\ and \ha\ fluxes remain unchanged,
because the wavelength regions determining the continua
and the ``\hd+\hc+\hb+\ha'' lines have much larger total weights
than that determining the abnormal Fe\,II multiplets.
We thus adopt the fits that used the uniform
factor for the whole \feii\ spectrum,
since we are not concerned with the properties of \feii\ emission
in this work.
Example spectra as well as the fitting results are  demonstrated in Fig. 1.

\subsection{Refined Fitting of Emission-line Profile} 
We re-calculate the reduced $\chi^2$ of the fits around \hb\
and the \ha\ regions (4750--5050\AA\ and 6400--6800\AA, respectively)
using the {\em original} errors for the 446 spectra.
Spectra having a fit with the reduced $\chi^2 > 1.1$ are picked up
for further refined line-profile fitting using the code described in detail
in Dong et al. (2005).
Briefly, we fit each pseudo-continuum subtracted spectrum
using various schemes and the one with the minimum reduced $\chi^2$ is adopted.
Broad lines are fitted with multiple Gaussians, as many as 4 at most;
the fits are accepted when the reduced $\chi^2 \leq 1.1$ \emph{or}
it cannot be improved significantly by adding in one more Gaussian (up to 4)
with a chance probability less than 0.05 according to F-test.\footnote{
We have found through our experiment that these criteria based on
$\chi^2$-test and F-test work well, although, theoretically,
these goodness-of-fit tests holds only for linear models (cf. Lupton 1993).}
Narrow Balmer lines mostly have similar profiles to \nii, \sii\ or the line core
of [O\,III] $\lambda5007$.
At this stage, the broad Balmer lines are not required to have the same profiles.
Fig. 2 shows examples of refined line-profile fitting in the \hb\ and \ha\ regions.
The line fluxes of \hb\ and \ha\ are listed in Table 1, and
the broad-line profile parameters are listed in Table 2.
The data and the fitted spectral parameters are available online
for the decomposed spectral components (continuum, \feii\ and other emission lines)
for the 446 objects.\footnote{Available at \\
http://staff.ustc.edu.cn/\~{}xbdong/Data\_Release/blueAGN\_DR4/,
together with an auxiliary code to explain the parameters and to demonstrate the fitting.}

Our two-step procedure of pseudo-continuum and line-profile fitting
has proved to be robust and self-consistent.
As a reliability check, we perform the first-step fitting (as in \S2.3)
by setting the initial values of the free parameters of emission lines
(except Iron lines) to those yielded from the refined line-profile fitting,
and find that the best-fits of both the continuum
and the Fe\,II emission are almost unchanged.

\subsection{Estimation of Parameter Uncertainties} 

As we have argued above, estimation of the measurement errors of
emission-line fluxes is important for deriving the intrinsic distribution of
the Balmer decrements.
The errors of line fluxes provided by MPFIT are unreasonably small:
the median errors of broad \hd, \hc, \hb\ and \ha\
are 13\%, 12\%, 6\% and 4\%, respectively.
They do not account for the uncertainty
introduced by the pseudo-continuum subtraction that
severely complicates measurement of the broad Balmer lines
(see Marziani et al. 2003 for a detailed discussion).
To take this and other possible effects into account,
we adopt a bootstrap approach to estimate the typical errors
for the whole sample.
We generate 500 spectra by randomly combining
the scaled model emission lines of one object (denoted as `A')
to the emission-line subtracted spectrum of another object (denoted as `B').
The emission line model of object `A' is scaled in such a way that
it has the same broad \hb\ flux as object 'B',
in order to  minimize changes in S/N within the emission-line
spectral regions in the simulated spectra.
Then we fit the simulated spectra following the same
procedure as described in \S2.3 and \S2.4.
For each parameter, we consider the error
typical of our sample
to be the standard deviation of the relative
difference between the input ($x_i$) and the recovered ($x_o$),
$\frac{x_o - x_i}{x_i}$.
These relative differences turn out to be normally distributed
for every parameter.
The thus estimated typical 1$\sigma$ errors
for the fluxes of broad \hd, \hc, \hb\ and \ha\
are 19\%, 19\%, 8\% and 5\%, respectively.
Hence the measurement error of the \bha/\bhb\ ratio follows the
log-normal distribution with 1$\sigma$=0.04 dex.
The uncertainties for the broad \hd\ and \hc\ line fluxes
are large because they are relatively weak
and/or are often severely contaminated  by the Fe\,II
and [O\,III]$\lambda4363$ emission.
Considering the large uncertainties of their measurements,
we do not discuss the \bhd/\bhb\ and \bhc/\bhb\ ratios in this study.

The errors provided by MPFIT are unreasonably small
also for the power-law indices and normalizations,
commonly being only 0.5 and 0.4 per cent, respectively.
Using the above boot-strap approach, the typical errors are 8 per cent
for $\alpha_{\lambda,1}$ and $\alpha_{\lambda,2}$,
5 and 3 per cent for $c_1$ and $c_2$, respectively.
The mean errors of the fluxes of narrow \hb\ and \ha\
are 6 and 5 per cent, respectively,
which are almost unaffected by the pseudo-continuum subtraction.

\section{Results}
\subsection{Intrinsic broad line \ha/\hb\ Distribution}

The measured \bha/\bhb\ ratios range from 2.3 to 4.2,
with a mean of 3.1 and skewed towards the large ratio end.
Considering the log-normal distribution of their measurement errors,
we plot the histogram in base-10 logarithm form
by dividing the sample into 20 bins, as shown in Fig. 3.
The profile of the distribution is very similar to Gaussian.
We fit the distribution with a Gaussian function
by minimizing \chisq\ assuming Poissonian errors for the counts in each bin.
A good fit (see Fig. 3) is achieved with a minimum reduced $\chisq = 0.89$
(17 degrees of freedom),
yielding a mean of $0.486 \pm 0.002$
and a standard deviation of $0.046 \pm 0.002$.
This model distribution is
identical to the histogram at the 96.5\% significance level
according to the Kolmogorov-Smirnov test.
We re-fit the histogram by varying the number of bins
from 10 through 30 and find that
all the fits yield similar results at similar confidence levels.
Therefore, the observed distribution of the \bha/\bhb\ ratios
is well described by log-normal, with a peak at $ \bha/\bhb = 3.06$.
This fact, together with that the measurement errors have a
log-normal distribution (see \S2.5), indicate that the distribution of the
intrinsic \bha/\bhb\ ratios should be, at least very close to,
log-normal.
The {\em intrinsic} \bha/\bhb\ distribution can be approximated by
deconvolving the measurement uncertainty ($1\sigma \sim 0.04$ dex)
from the observed distribution,
yielding a standard deviation as small as about 0.03 dex.
The deconvolved intrinsic distribution is also displayed in Fig. 3.

It should be noted that measurement of the broad \ha\ and \hb\ line
fluxes is insensitive to the deblending with the narrow \ha\ and
\hb\ components because of the weakness of the latter in blue AGN,
which account for only $\sim$5 and $\sim$4 per cent of the total
\ha\ and \hb\ fluxes in the sample. The distribution of the total
line flux ratios $\log(\frac{\ha^{total}}{\hb^{total}})$ has a mean
of 0.493 and a standard deviation of 0.046, nearly identical to
those of the broad components $\log(\frac{\ha^{b}}{\hb^{b}})$.

\subsection{Dependence of \bha/\bhb\ on AGN Properties}
We conduct comprehensive statistical analysis using the present
sample in an attempt to investigate whether there exist any
systematic trends (\textit{biases}) of the \bha/\bhb\ dependence on
AGN types and/or properties, such as radio-loud objects, objects
with clumpy/double-peaked line profiles, objects in low states. This is
important as any biases, if being significant, would affect the
observed distribution of \bha/\bhb. Correlation analysis is
performed between \bha/\bhb\ and various AGN properties, and the
results are summarized in Table 3. We report the
Spearman rank correlation coefficient ($\rho$) and the probability
($P_{\rm null}$) that a correlation is not present. When upper limits
are present, we use the generalized Spearman rank correlation test
as implemented in the ASURV package (Isobe et al. 1986). We also
perform statistical tests to compare the Balmer decrements among
various subsamples.

\subsubsection{Luminosity and Eddington Ratio}
We calculate monochromatic continuum luminosity
$\lfive \equiv \lambda L_{\lambda}$ at 5100\AA\ from the
power-law fit $a_1 \lambda ^{-\alpha_{\lambda,1}}$.
Due to the blue slope criterion the optical luminosities
have little contribution from host galaxy starlight.
Based on the empirical BLR radius--luminosity (R--L) relationship and
the assumption that the BLR gas is virialized under the control of gravity of
the central supermassive black hole,
black hole mass $M_{BH}$ can be estimated using the so-called
linewidth--luminosity--mass scaling relation (e.g., Kaspi et al. 2000).
We calculate the black hole masses using
the formalism presented in Vestergaard \& Peterson (2006, their equation 5)
with \hb\ FWHM measured from the best-fit model of \hb\ broad-line profiles.
This formalism was calibrated with reverberation mapping-based masses and
used the R--L relation by Bentz et al. (2006) corrected for host galaxy
starlight contamination.
The Eddington ratios (\lratio) are calculated assuming
a bolometric luminosity correction $L_{bol} \approx 9\lfive$
as for normal QSOs (Kaspi et al. 2000, Elvis et al. 1994).
No correlation between the Balmer decrement and \lfive, \lratio\ and $M_{BH}$
are found to be significant at the $P_{null} <0.01$ level (see Table 3).
To check whether the results are dependent on the exact formalisms
for black hole mass estimation,
we re-calculate black hole masses using various formalisms, including
Peterson et al. (2004) and Onken et al. (2004) and the R--L relation
from Kaspi et al. (2005) and Bentz et al. (2006), respectively.
For the line width, we also use the second moment of the line profile
($\sigma_{line}$, often referred to as ``line dispersion'')
measured from the model line profile.
To avoid possible contamination to the continuum luminosity from jet
and host-galaxy starlight, we also try to use the \hb\ luminosity to
calculate the black hole masses.
Yet no correlations are found, either.
We also consider the possibility that these
correlations may be present but differ for different black hole
masses, and as such the trend of the correlations could be reduced
when the sample spans a large range of black hole masses.
We divide the sample into several log\mbh\ bins with a bin size of
0.4 dex for each bin. We find that, in none of the bins,
the correlations of \bha/\bhb\ with \lfive\ or \lratio\ are present.
The relations of \bha/\bhb\ with \lfive\ and \lratio\ are displayed in
Fig. 4a and 4b, respectively, where the 122 objects in the \mbh\ bin
of $10^{7.8} -10^{8.2}$ \msun\ are denoted as solid black squares.

The luminosity \lfive\ ranges from $2\times 10^{43}$ to
$2\times 10^{45}$ \lum\ with a median of $2\times 10^{44}$ \lum.
We compile two subsamples with $L_{5100} > 5\times10^{44}$ \lum\
(`high-L' sample, 45 objects) and
$L_{5100} < 1\times10^{44}$ \lum\ (`low-L' sample, 49 objects), respectively.
The mean and standard deviation of $\log(\frac{\ha^{b}}{\hb^{b}})$
in the `high-L' subsample are 0.496 and 0.046, respectively;
and 0.479 and 0.049 respectively in the `low-L' subsample.
The mean values of the two samples are consistent with each other,
with a chance probability of 9 per cent
according to the Student's \textit{t}-test.
The Eddington ratios range from 0.005 to 1.2 with a median of 0.2.
We also compile two subsamples of objects with
$\lratio > 0.5$ (`high-state', 57 objects) and
$\lratio < 0.03$ (`low-state', 31 objects), respectively,
which have the mean and standard deviation of
$\log(\frac{\ha^{b}}{\hb^{b}})$ of 0.488 and 0.043,
and 0.499 and 0.058, respectively.
Again, the two subsamples have mutually consistent mean values,
at a chance probability of 35 per cent given by Student's \textit{t}-test.

\subsubsection{Radio loudness}
We cross-correlate the sample with the Faint Images of the Radio Sky
at Twenty cm (FIRST; Becker et al. 1995) Survey following the procedure
described in Lu et al. (2007).
For unresolved FIRST sources a matching radius of 3\arcsec\ is used;
for resolved sources the FIRST images are visually inspected.
In this way 68 matches are obtained.
For these matches, we define and calculate the
radio-to-optical flux ratio (radio-loudness)
following Ivezi\'c et al (2002),
$R_i \equiv \log (\frac{f_{20cm}}{f_i})$,
where $f_i$ and $f_{20cm}$ are the flux
densities at \textit{i}-band and 20~cm, respectively.
For the rest 336 objects that were covered by FIRST,
we calculate the upper limit of $R_i$ by taking the FIRST detection
limit of 1 mJy as the flux limits at 20 cm.
Fig. 4c shows the \bha/\bhb\ ratio versus $R_i$.
The correlation is weak at most, as indicated by $\rho = 0.205$
incorporating both detections and upper limits of $R_i$.
We also calculate the k-corrected radio power at 20 cm as
$P_{20cm} = 4 \pi D_L^2 f_{int} /(1+z)^{1+\alpha_r} $,
where the radio spectral index
$\alpha_r$ ($F_\nu \propto \nu^{\alpha_r}$)
is assumed to be $-0.5$ for all the objects.
Similarly, there seems at most a weak correlation,
as indicated by $\rho = 0.209$.

It is well known that the radio-loudness distribution appears to be
bimodal\footnote{However, the genuineness of this bimodality is
still a matter of debate; see, e.g., Hooper et al. (1995), White et al. (2000)
and Cirasuolo et al. (2003).}
with $\sim$90\% being radio-quiet (RQ) and $\sim$10\% radio-loud (RL)
(Kellermann et al. 1989, Ivezi\'c et al 2002 and references therein).
Hence for the two populations the central engines or physical processes
related to accretion/jet may be different
(e.g., Blandford et al. 1990, Boroson 2002, Falcke et al. 1996).
To further examine whether the Balmer decrements are systematically distinct
for the two populations, we compile a RL and a RQ subsample
following Ivezi\'c et al (2002):
19 sources with $R_i > 1$ are regarded as RL,
while the rest 385 sources covered by FIRST are RQ.
The mean and standard deviation of $\log(\frac{\ha^{b}}{\hb^{b}})$
in the RL subsample are 0.528 and 0.057, respectively;
and in the RQ subsample they are 0.483 and 0.046, respectively.
The mean Balmer decrement appears to be slightly larger in the RL subsample
by 0.05 dex than in the RQ one.
This difference is significant according to Student's \textit{t}-test,
with a chance probability $\ll 0.01$.

\subsubsection{Line Profile}
It has been known for a long time that the Balmer decrement is
larger in the line core than in the line wings
(e.g., Shuder 1982, Crenshaw 1986, Stirpe 1991,
Korista \& Goad 2004 and references therein).
As a result the integrated Balmer decrement might depend
on the line profile to some extent.
To investigate this issue, we compute the skewness (the 3rd moment)
and kurtosis (the 4th moment)
based on the best-fit model of \hb\ broad-line profiles,
as well as FWHM and $\sigma_{line}$ obtained in \S3.2.1.
The correlation between the integrated Balmer decrement and skewness
is at most weak, if exist at all, as indicated by $\rho = -0.203$.
No correlations are found between the integrated Balmer decrement and
kurtosis, FWHM and $\sigma_{\rm line}$, respectively.
Considering that skewness and kurtosis are sensitive to the errors
in the line wings caused by the substraction of the Fe\,II, [O III]
and He\,II emission, we also compute some empirical yet robust profile
parameters based on the model profile.
We define three indices to characterize
the asymmetry (\textit{viz.} skewness, $AI$), shape ($SI$) and kurtosis ($KI1$) as,
$AI =(C(\frac{3}{4}) - C(\frac{1}{4}))/$FWHM,
$SI =(FW(\frac{1}{4}) + FW(\frac{3}{4}))/(2\times {\rm FWHM})$,
and $KI1 =FW(\frac{3}{4})/FW(\frac{1}{4})$,
following De Robertis (1985), Boroson \& Green (1992)
and Marziani et al. (1996), respectively.
Where $C(\frac{1}{4}), C(\frac{3}{4})$ is the centroid at
$\frac{1}{4}, \frac{3}{4}$ maximum,
and $FW(\frac{1}{4}), FW(\frac{3}{4})$,
the full width at $\frac{1}{4}, \frac{3}{4}$ maximum, respectively.
These three dimensionless parameters thus defined are not affected
by the choice of the rest frame.
Collin et al. (2006) characterize the broad-line profiles
by the ratio of FWHM to $\sigma_{line}$,
based on which the broad-line profiles of AGN are separated into two categories:
the first, having $\frac{FWHM}{\sigma_{line}}<2.35$,
are narrower lines with relatively extended wings;
the second, having $\frac{FWHM}{\sigma_{line}} \geq 2.35$,
are broader lines being relatively flat-topped.
This linewidth ratio is similar to the above kurtosis index ($KI1$),
and is therefore also computed and denoted as $KI2$.
We perform correlation analysis of the integrated Balmer decrement
with the above various parameters, and find no significant corrections
(see Table 3).
The relation between the integrated Balmer decrement and $KI1$ is
displayed in Fig. 4d.

Among our blue AGN sample, there are 23 objects having \ha\ or \hb\
lines of double or even multiple peaked profiles\footnote{
Such profiles are defined as having the number of peak and
``pseudo-peak'' greater than 2;
a ``pseudo-peak'' is defined as the point where the 2nd derivative
is minimal and negative. See Shang et al. (in preparation) for details.}
(Shang et al., in preparation), with 5 being radio-loud.
Such profiles are found in about 3 per cent of AGN (Strateva et al. 2003)
and have a higher occurrence in LINERs (Eracleous 2004)
and in RL sources ($\sim$20\%, Eracleous \& Halpern 2003).
Using a sample selected from RL AGN, Eracleous \& Halpern (1994)
found that double-peaked emitters have the large integrated Balmer decrements,
like their parent population of RL AGN;
later they further found that their integrated Balmer decrements are
even larger than that of the latter
(on average 5.23 versus 4.26, Eracleous \& Halpern 2003).
We investigate this issue using the above 23 objects (DBP subsample).
The mean and standard deviation of $\log(\frac{\ha^{b}}{\hb^{b}})$
in the DBP subsample are 0.515, 0.055, respectively,
while these values are 0.484 and 0.046 in the rest of objects (non-DBP subsample).
The mean integrated Balmer decrement appears to be slightly larger
in the DBP subsample than in the non-DBP sample,
which is significant with a chance probability of 0.002 (0.01)
by Student's \textit{t}-test assuming the two distributions to have
the same (different) variance.
We cannot find any difference in integrated Balmer decrement
between the DBP subsample and the RL subsample
(with a \textit{t}-test probability of 44\%),
possibly due to the small size of the two subsamples.

\subsubsection{Other SED properties}
The shape of the ultraviolet to X-ray ionizing continuum affects the
extended partially ionized zone and thus the Balmer decrements (Kwan
\& Krolik 1981). Here we take the \aox\ values
(the slope of a hypothetical power law between 2500\AA\ and 2 keV)
for 268 matched sources from Anderson et al. (2007).
These values were derived from broadband X-ray (ROSAT) and
\textit{g}-band (SDSS) fluxes by assuming
an X-ray energy index $\alpha_{\rm x} = 1.5$
and an optical index $\alpha_{\lambda} = 1.5$.
No correlation between the Balmer decrement and \aox\ is found.
We also perform correlation analysis between the Balmer decrement and
the optical--near-ultraviolet continuum slope ($\alpha_{\lambda,1}$);
and find no significant correlation (see Table 3).
The relations of the \bha/\bhb\ ratio with \aox\ and $\alpha_{\lambda,1}$
are displayed in Fig. 4e and 4f, respectively.

\section{Discussion}
\subsection{Representativeness of Unreddened AGN}

Internal Dust extinction in objects of our blue AGN sample should be negligible.
In the sample, $\alpha_{\lambda,1}$ varies in the range
from 1.5 to 2.7 (see Table 1) with a standard deviation of 0.2.
Such a scatter is consistent with previous results that
there is a large {\em intrinsic} dispersion of the continuum slope
(Elvis et al. 1994, Rowan-Robinson 1995, Natali et al. 1998, Kuhn et al. 2001).
To check how much the sample suffers from extragalactic dust extinction,
we cross-match these AGN with the \textit{Galaxy Evolution Explorer}
(GALEX; Morrissey et al. 2005) General Data Release 3
with a matching radius of 2.\arcsec6 in a way similar to Trammell et al. (2007).
We obtain 253 objects with both reliable far-ultraviolet ($f$)
and near-ultraviolet ($n$) magnitudes,
while 185 of the rest have not been covered by GALEX yet.
For the 253 objects we compute the relative $f-n$ colors defined as
the $f-n$ colors minus the corresponding median color of quasars
at the same redshift. The digital curve of median color versus redshift
is kindly provided to us by G.\,Trammell,
as presented in Fig. 13 of Trammell et al. (2007).
We find the distribution of the relative $f-n$ colors can be well described
by a Gaussian with a mean of zero and a standard deviation of 0.3,
a dispersion caused by GALEX photometric uncertainties only
(Trammell et al. 2007, see their Fig. 14).
We also find that neither the $f-n$ color nor the relative $f-n$ color
correlates with the Balmer decrement. Hence we believe that there is
little dust extragalactic extinction in our sample.

We notice that there exists a potential bias against selecting AGN
with intrinsic red slopes by our criterion.
To test the representativeness of our sample,
we perform Spearman correlation analysis of the continuum slope
with all the various, above-mentioned parameters in the same way as in \S3.2.
The results are listed in Table 3.
It turns out that no correlations of $\alpha_{\lambda,1}$ are
present with the Eddington ratios, black hole mass, radio-loudness,
radio power, linewidth and various other line-profile parameters.
But there is a positive correlation between the continuum slope and
the nuclear luminosity \lfive\ as indicated by
$\rho = 0.457$ and $P_{null} = 10^{-6}$ (see Table 3).\footnote{
Such a slope--luminosity correlation has also been reported in the literature
(e.g., O'Brien et al. 1988, Cheng et al. 1991, Francis 1993,
Carballo et al. 1999, Kuhn et al. 2001).
Previously, however, it was not clear whether the slope--luminosity
correlation is intrinsic or just arises from the correlation between
the slope and redshift in a magnitude-limited sample.
Some authors claimed an intrinsic dependence of the continuum slope
with redshift rather than with luminosity
(O'Brien et al. 1988, Cheng et al. 1991, Francis 1993).
Our sample is much larger than the previous ones and
has a limited redshift range $z<0.35$.
In our sample no significant correlation between the continuum slope
and redshift is found;
yet the significant correlation between slope and \lfive\ still exists
by partial correlation analysis under the control of redshift,
with a coefficient of 0.36 and a chance probability $\ll 10^{-4}$.
As a by-product of this study, we suggest that the slope--luminosity
correlation is intrinsic.}
Thus our sample is biased toward relatively high luminosity.
This is consistent with the fact that the sample covers a high luminosity range
as $\lfive \gtrsim 2 \times 10^{43}$ and
\ha\ luminosity $\gtrsim 5 \times 10^{41}$ \lum.
To sum up, our sample is free from biases regarding the
above-mentioned AGN properties except luminosity.

\subsection{On the Intrinsic \bha/\bhb\ Distribution and Its Applications}
As discussed in \S1, historically the broad-line Balmer decrements
in Seyfert 1s and QSOs have been generally thought to be
considerably steeper than the Case B value \emph{and} to have a
large intrinsic dispersion. Such a belief may have originated from
results of early studies with small samples,
and was (mis-)reinforced by observations of Seyfert 1.8/1.9 and
broad-line radio galaxies,
and further assured by its consistence with photoionization modeling
mainly of individual clouds in various conditions
(see observational and theoretical references in \S1).
In a sample of 36 Seyfert 1s (including Seyfert 1.2 and 1.5) studied
by Osterbrock (1977), the mean \ha/\hb\ ratio is 3.6 with a range
from 2.6 to 5.9 (uncorrected Galactic reddening).
Similar results were found in a sample of 24 Seyfert 1/QSOs
by Neugebauer et al. (1979),
giving the mean \ha/\hb\ ratio 3.6 with a range from 2.2 to 4.9
(uncorrected Galactic reddening; \ha\ measurements include \nii\ flux).
In addition to relatively large errors inherent in the early measurements and
the neglect of Galactic reddening correction,
we believe that internal reddening also plays a non-negligible role
in causing the steeper Balmer decrements in previous results.
Similar situation was found in PG quasars
(Rowan-Robinson 1995, Baskin \& Laor 2004).
For instance, I\,ZW\,1 has an observed \ha/\hb\ ratio of 4.86
as given in Table 1 of Osterbrock (1977),
but actually its broad lines suffer internal reddening of
$E(B-V) \sim 0.1$ (Rudy et al. 2000)
as well as Galactic reddening of $E(B-V) \sim 0.1$
(Schlegel et al. 1998; Stark et al. 1992).
Interestingly, for the 9 Seyfert 1s listed in Table 2 of Wu et al. (1983),
the mean \bha/\bhb\ ratio is 3.1 after individual correction for Galactic
reddening (Goodrich et al. 1990), which is almost the same as ours found here.
Recently, in a sample of broad-line AGN with low starlight contamination,
Greene \& Ho (2005) found that the mean value of the
total (narrow + broad) \ha/\hb\ ratio is 3.5;
the somewhat larger value compared to ours should also be due to
internal reddening. In a large sample of about 2000 narrow-line Seyfert 1s (NLS1s),
Zhou et al. (2006) found the mean \bha/\bhb\ ratio to be 3.0,
close to the value derived here for blue AGN.
In summary, results obtained from early small samples and from
recent much large samples all point to that the mean \bha/\bhb\ ratio
of Seyfert 1/QSOs is only slightly larger than the Case B value.

A surprising feature emerged from the present study is the actual
dispersion of the intrinsic \bha/\bhb\ ratio being rather small
(0.03 dex in a log-normal distribution)!
This is contrary to the prevalent belief that
there is a ``considerable range of intrinsic (Balmer) line ratios''
(Wu et al. 1980), which was based on early samples that were in fact
too small to make a general conclusion.
Our conclusion should hold for luminous Seyfert 1s and QSOs in general,
since by selection our sample is only biased against AGN
with intrinsic red continua, but the Balmer decrement does not
correlate with the continuum slope as found in \S3.
We further suggest that our result is also likely to hold
for AGN of low luminosity.
Firstly, there is no correlation between the Balmer decrement and luminosity
in our sample, and the `high-L' and `low-L' subsamples
have the indistinguishable mean Balmer decrements.
Secondly, in the large sample of about 2000 NLS1s that are generally
at high accretion states, the Balmer decrements still cluster
tightly around $\sim$3 even when the nuclear \lfive\ goes down to
$10^{41}$ \lum\ (Zhou et al. 2006).
Moreover, as shown in \S3, the Balmer decrement does not correlate
with the Eddington ratio.
Recently, contrary to our findings here,
La~Mura et al. (2007) found a weak correlation between
the \bha/\bhb\ ratio and the \lratio.
However, as suggested by those authors,
this might result from the inclusion of reddened objects in their sample.
They also found a weak correlation between
the flux ratios of broad-line Balmer series (up to \hd)
and the line width;
but, in the case of \bha/\bhb\ ratio,
the correlation is not significant as indicated by
$P_{\rm null} = 3.34 \times 10^{-2}$.

It should be noted that in some variable objects
an anti-correlation between the continuum flux (i.e., accretion state)
and the \bha/\bhb\ ratio has been reported
during flux variability over a time scale of months
(e.g., the prototypal NGC\,5548; Wamsteker et al. 1990,
Dietrich et al. 1993, Shapovalova et al. 2004).
Such a behavior has been well predicted and explained by theoretical models
(e.g., Netzer 1975; Korista \& Goad 2004).
Thus the lack of such a correlation in the AGN ensemble seems to be
somewhat unexpected.
We guess that, instead of instantaneously responding to the continuum variation,
the BLR clouds distribution (as a function of mass and/or luminosity)
is perhaps determined/adjusted by the long-term average of the accretion state;
for an AGN ensemble, this long-term adjustment may make the average
\bha/\bhb\ ratio in a narrow range.
In addition, according to Table 8 of Shapovalova et al. (2004)
and Table 2 of Dumont et al. (1998),
the Balmer decrement of NGC\,5548 varies between 3.0 and 4.3,
which is not extreme and is actually within the range
for the blue AGN ensemble as reported here.
Due to the small variability amplitude,
any correlation, even if exists in individual objects,
would be smeared out in the ensemble.
In a few cases, such kind of variability can be explained
by variation in extinction
(Goodrich 1989, Tran et al. 1992, Goodrich 1995;
see a discussion about the time scale of this kind of variability in Wang et al. 2007).

From theoretical perspective, however, a tight \bha/\bhb\
distribution around 3.1 seems to be quite surprising---all
the BLR photoionization computations predict a rather large range
in the Balmer decrement for individual clouds in plausible BLR conditions
(see references in \S1).
It seems unlikely that such a discrepancy is mainly caused by the
incapability of photoionization modeling of the BLR clouds, because
all the model computations over the past 40 years gave the
similar trend of large \bha/\bhb\ range
(e.g., Netzer 1995, Dumont et al. 1998, Korista \& Goad 2004),
although such computations for a single cloud are still uncertain
(see Netzer 1995 and references therein).
Another possibility, in fact an old proposal, is that
the BLR emitting gas of all AGN have been ``fine-tuned''
to a certain ionization parameter by some physical processes
(e.g., the ``hot-warm'' model of Krolik et al. 1981).
However, the existence of such a fine-tuning,
among clouds in likely jumbled environments as in AGN BLRs,
is questioned by others and deemed to be unnatural and unlikely
(e.g., Mathews \& Ferland 1987; Baldwin et al. 1995).
Here, we suggest a plausible explanation to this seemingly discrepancy
invoking the ``locally optimally-emitting cloud'' (LOC)
model, as proposed by Baldwin et al. (1995).
The essential idea of the LOC model is that each line arises
predominantly from clouds only in a narrow range of density and
distance from the continuum source, due to natural selection effects
largely introduced by the atomic physics.
In this scenario, the similar values of the Balmer decrements,
just like other surprising similarities in emission-line spectra of
Seyfert 1/QSOs (Davidson \& Netzer 1979, Baldwin et al. 1995),
appear to be a natural consequence.
According to the recent calculation by Korista \& Goad (2004, their Figure 5)
using the spectral synthesis code CLOUDY (version 90.04),
the \bha/\bhb\ flux ratio varies from roughly 17 to approximately 1
across the parameter plane of gas density and ionizing flux
characteristic of BLR clouds;
the \bha/\bhb\ ratio, when integrated over the full BLR of their LOC model,
varies from 3.7 (high state) through 4.9 (low state).
Using CLOUDY version 07.02 (last described by Ferland et~al.\ 1998)
with improved collisional rates for excited states of hydrogen atoms,
the \bha/\bhb\ ratio contours now have values that are typically
$\sim$0.1 dex {\em smaller} than those presented in Figure~5 of
Korista \& Goad (2004) (Korista, private communication).
Thus there exists a large span of cloud parameters within the density-flux plane,
over which most of the broad emission lines are emitted,
for which the \bha/\bhb\ ratios lie between 2.5 and 4.
In fact LOC integrations over the range of cloud parameters
adopted by Korista \& Goad (2004) now predicts
an \bha/\bhb\ ratio that is consistent with our measurements
(Korista, private communication).
We therefore suggest that the LOC model,
together with improvement on the photoionization modeling,
can give a natural explanation to the seemingly discrepancy
between the theory and the measurements.

As is found here, radio-loud objects and objects having
double-peaked emission-line profiles among the blue AGN ensemble
have slightly larger Balmer decrements than the rest on
average. In these kinds of AGN the physical conditions related
to accretion/jet may be different
(Blandford et al. 1990, Falcke et al. 1996, Eracleous \& Halpern 2003),
e.g., with additional X-ray ionizing radiation from the jet base.
Thus collisional excitation, self-absorption, and other line-transfer effects
may play roles to enlarge the Balmer decrements, as discussed by some authors
(e.g., Osterbrock et al. 1976, Crenshaw et al. 1988, Netzer et al. 1995).
Given the small fractions of these two classes
($\sim10$ and 3 per cent, respectively) among the AGN population,
the distribution of the \bha/\bhb\ ratios show little changes
when the RL and double-peaked objects are excluded.
In another word, for the bulk of the AGN population,
such effects are insignificant.

Our finding out of this study may have an interesting
implication: the precise distribution of the intrinsic \bha/\bhb\
ratio, that is insensitive to any nuclear properties known so far,
renders it a useful tool with which the BLR extinction can be
derived, at least in a {\em statistical} manner.
Specifically, for a sample concerned, we can derive its
internal $E(B-V)$ distribution by de-convolving
the {\em observed} \bha/\bhb\ distribution with
the {\em intrinsic} \bha/\bhb\ distribution as found here.
This approach has now been applied to a large sample of broad-line AGN
culled from the SDSS DR4 to derive the internal $E(B-V)$ distribution
of the AGN BLR in the local universe, and hence to obtain the
fraction of obscured AGN of various intrinsic luminosity and of
various degree of extinction (Zhang et al., in preparation).
At the zeroth-order approximation, Dong et al. (2005) have used
the Balmer decrement to derive the BLR reddening
for a sample of luminous Seyfert 1.8/1.9 galaxies culled from the SDSS EDR
and derived the fraction of partially obscured quasars in the local universe.
Follow-up XMM-Newton observations confirm that these sources have
large absorption column densities in the X-ray (Zhou et al., in preparation).

\section{Summary}
We have investigated the broad-line Balmer decrements for a large,
homogeneous sample of 446 blue AGN, of Seyfert 1 galaxies and QSOs.
They are selected from the Sloan Digital Sky Survey Fourth Data
Release according to the criteria of redshift $z \lesssim 0.35$, the
median spectral signal-to-noise per pixel $\geq 10$, and the
continuum slopes $\indw \gtrsim 1.5$ ($f_{\lambda} = \lambda ^{-\indw}$)
that are fitted in the rest-wavelength range of 4000--5600 \AA.
With the blue criterion of the continuum slope,
dust extinction in the sample objects is expected to be negligible,
which is also confirmed by their relative colors in the ultraviolet.
The sample is fairly representative of normal Seyfert 1/QSOs
(at least the luminous objects with \ha\ luminosity greater than $10^{41}$ \lum),
in light of the fact that the optical--near-ultraviolet continuum slope
does not correlate with the Balmer decrement,
nor with other AGN properties except nuclear luminosity.
We find that (i) The distribution of the intrinsic broad-line
\ha/\hb\ ratios can be well described by log-normal, with a peak at
\ha/\hb\ =3.06 and an dispersion of likely 0.03 dex only. (ii) there
are no significant corrections between the Balmer decrement and the
nuclear properties such as luminosity, accretion rate, continuum
slope and $\alpha_{OX}$; (iii) on average, the Balmer decrements are
slightly larger in radio-loud objects (3.37) and
objects having double-peaked emission-line profiles (3.27).
Therefore we suggest that the broad-line \ha/\hb\ ratios
can be used as a good indicator of dust extinction of the AGN broad-line
regions, at least in a {\em statistical} manner. This result is
especially true for radio-quiet AGN with regular emission-line profiles
that constitute the vast majority of the AGN population.
Such an application has significant implications for deriving the
distribution of internal dust extinction in the BLR of AGN, and
hence the obscuration fraction of AGN in the universe.

\section*{Acknowledgements}
We thank the referee, Kirk Korista, for his enlightening suggestions
and comments which improved this paper significantly.
We thank G.~Trammell for providing us with the digital median color curves,
and thank J.~Krolik for his comments during and after the USTC AGN summer school.
Funding for the Sloan Digital Sky Survey (SDSS) has been provided
by the Alfred P. Sloan Foundation, the Participating Institutions,
the National Aeronautics and Space Administration, the
National Science Foundation, the U.S. Department of Energy, the
Japanese Monbukagakusho, and the Max Planck Society. The SDSS is
managed by the Astrophysical Research Consortium (ARC) for the
Participating Institutions. The SDSS Web site is
http://www.sdss.org/.
This work is supported by Chinese NSF grants
NSF-10533050 and NSF-10573015, the Knowledge Innovation Program
(Grant No. KJCX2-YW-T05) and the BaiRenJiHua project (W.~Yuan)
of the Chinese Academy of Sciences.
XBD is partially supported by a postdoctoral grant
from Wang Kuan-Cheng Foundation.

\clearpage
\begin{figure*}
\label{fig-1}
\includegraphics[width=16cm]{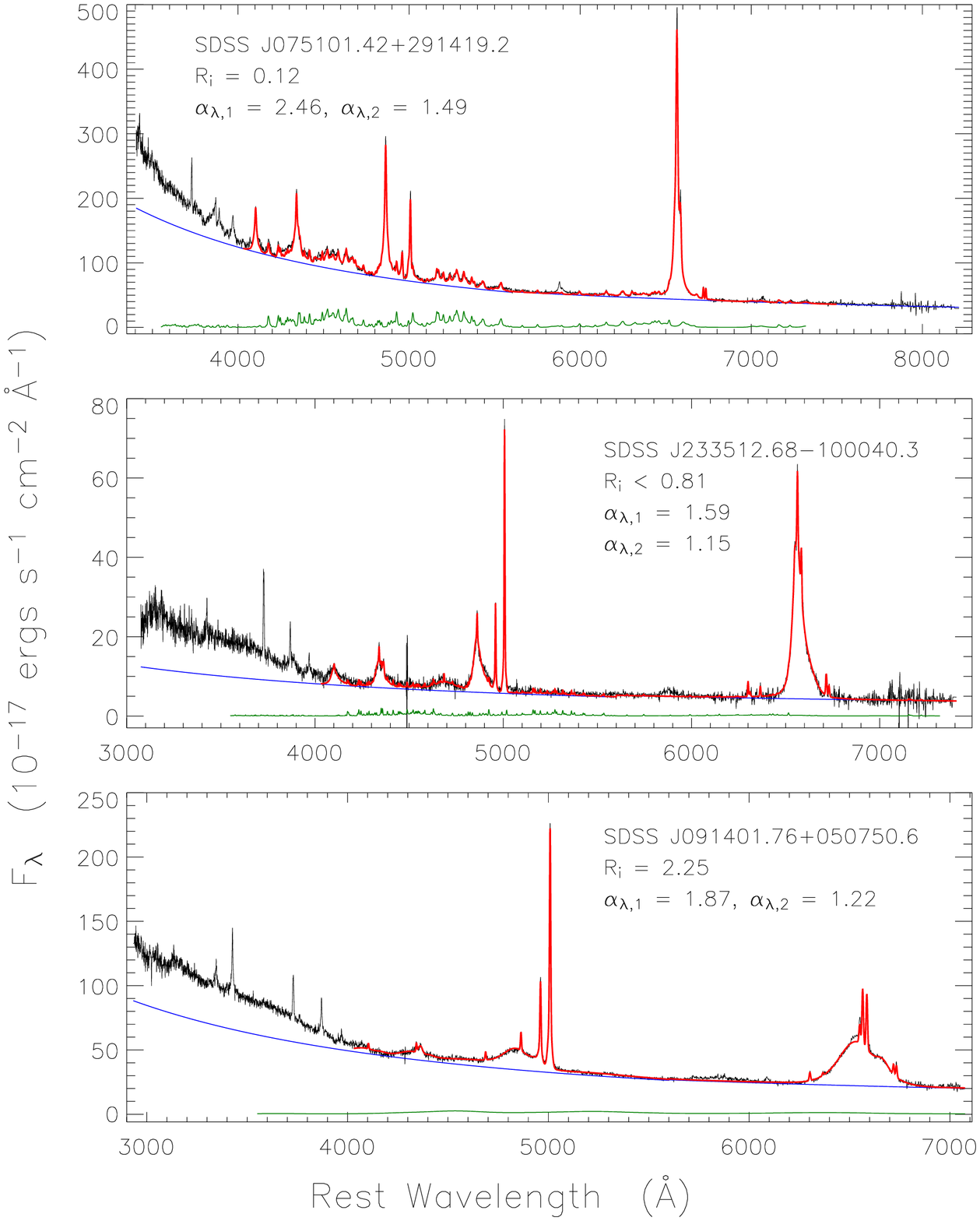}
\caption{Representative examples of the SDSS spectra and simultaneous fitting
of the continuum, Fe\,II and other emission lines from 4030\AA\ to 7500\AA.
In each panel, we plot the observed spectrum (black), the sum of the
best-fit components (red), the continuum modeled as a broken
power-law with a break at 5600\AA\ (blue), and the Fe\,II emission (green).
The radio-loudness ($R_i = \log (\frac{f_{\mathrm 20cm}}{f_i})$),
the continuum slopes blueward ($\alpha_{\lambda,1}$) and redward
($\alpha_{\lambda,2}$) of 5600\AA\ are indicated.
Note that the model Balmer broad-line profile in some objects
differ more or less from the observed one due to that the Balmer lines are
assumed to have the same profile in the fit; for such objects we
refit the line profiles to obtain much accurate line parameters
(see the text in \S2.4 and Fig. 2). }
\end{figure*}

\clearpage
\begin{figure*}
\label{fig-2}
\includegraphics[width=16cm]{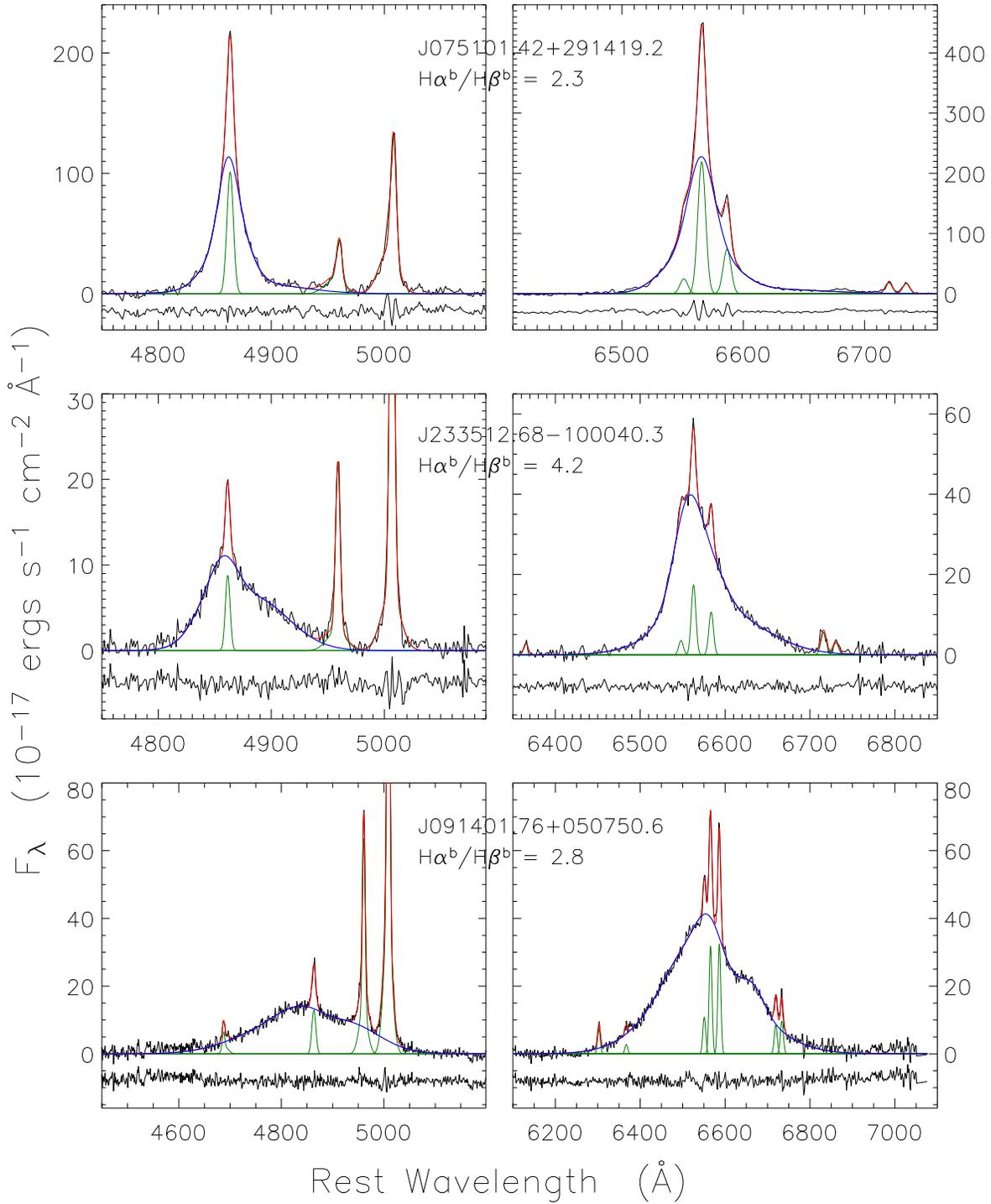}
\caption{Results of our line-profile fitting procedure
applied to the H$\beta$ region (left panels)
and the H$\alpha$ region (right panels)
for the 3 objects demonstrated in Fig. 1.
We plot the original data (black), the sum of all the best-fit components (red),
the fitted narrow lines (green), the fitted broad H$\beta$ and H$\alpha$ (blue),
and the residuals of the fit (bottom, offset downward for clarity).
The broad-line Balmer decrements are indicated. }
\end{figure*}

\clearpage
\begin{figure*}
\label{fig-3}
\includegraphics[width=8cm]{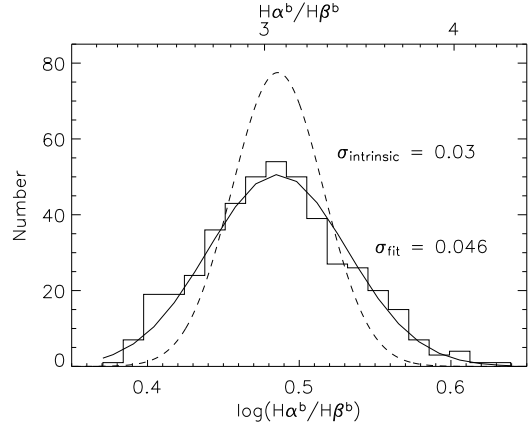}
\caption{Histogram of $\log(\frac{\bha}{\bhb})$ for the 446 blue AGN.
Also displayed are the fit (solid line) with a Gaussian function
yielding a mean of $0.486 \pm 0.002$ and a standard deviation of
$0.046 \pm 0.002$.
The intrinsic $\log(\frac{\bha}{\bhb})$ distribution is over-plotted
(dash line) with a standard deviation of 0.03, that is estimated by
de-convolving the observed distribution with the dispersion caused
by measurement uncertainty. }
\end{figure*}

\begin{figure*}
\label{fig-4}
\includegraphics[width=15cm]{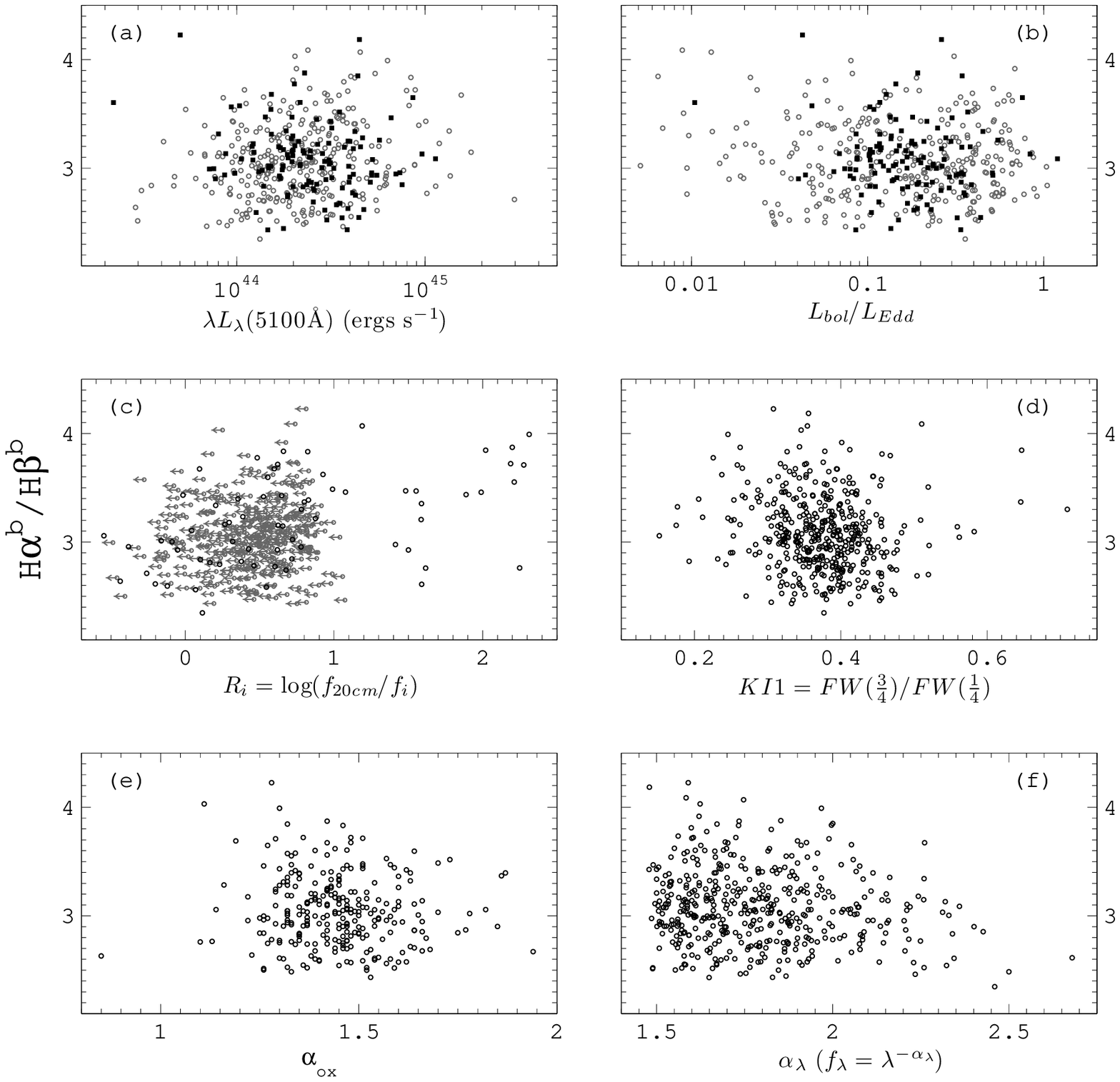}
\caption{Plots of \bha/\bhb\ versus $\lambda L_{\lambda}$(5100\AA),
\lratio, radio loudness $R_i$, kurtosis index $KI1$
as defined by Marziani et al. (1996), \aox,
and the optical--near-ultraviolet continuum slope $\alpha_{\lambda}$.
In panel (a) and (b), objects in the \mbh\ bin of $10^{7.8} -10^{8.2}$ \msun\
are denoted as solid black squares.
In panel (c), the 404 objects covered by FIRST are plotted, among
which 336 have only upper limits (grey circles with arrows).
In panel (e), the 268 sources having ROSAT matches are plotted, where
the \aox\ values are taken from Anderson et al. (2007).
Note that although the mean Balmer decrements are slightly larger
in the radio-loud subsample and the subsample of objects
with double-peaked line-profiles,
there are no significant correlations of the Balmer decrement
with $R_i$ or $KI1$ in the blue AGN ensemble. }
\end{figure*}


\clearpage
\begin{deluxetable}{rrrrrrrrrrrrrr}
\tablenum{1} \tablewidth{0pt} \topmargin 0.0cm \evensidemargin = 0mm
\oddsidemargin = 0mm
\tabletypesize{\tiny} \tablecaption{Properties of Blue AGN}
\tablehead{ \colhead{Object} & \colhead{$z$} & \colhead{$L_{5100}$}&
\colhead{F(H$\beta^{n}$)\tablenotemark{~a}} &
\colhead{F(H$\beta^{b}$)\tablenotemark{~a}} &
\colhead{F(H$\alpha^{n}$)\tablenotemark{~a}} &
\colhead{F(H$\alpha^{b}$)\tablenotemark{~a}} &
\colhead{$\alpha_{\lambda,1}$} & \colhead{$\alpha_{\lambda,2}$} &
\colhead{$f$} & \colhead{$n$} & \colhead{$i$} &
\colhead{$E_{B-V}^{Gal}$} & \colhead{$f_{20cm}^{int}$} \\
\colhead{(1)} & \colhead{(2)} & \colhead{(3)} & \colhead{(4)} &
\colhead{(5)} & \colhead{(6)} & \colhead{(7)} & \colhead{(8)} &
\colhead{(9)}& \colhead{(10)} & \colhead{(11)}& \colhead{(12)}&
\colhead{(13)} & \colhead{(14)} }
\startdata
000710.01$+$005329.0 & 0.31620 &  3.85 &   27.8 &  2778.8 &  146.0 &  9494.1 & 1.83 & 1.83 & 18.24 & 17.66 & 17.39 & 0.032 &   1.44   \\
000834.72$+$003156.1 & 0.26303 &  2.71 &  191.4 &  1002.1 &  574.2 &  2590.6 & 2.02 & 1.61 & 18.46 & 17.95 & 17.48 & 0.040 &   $<$1   \\
000943.14$-$090839.1 & 0.20958 &  2.09 &    6.7 &  2932.7 &   41.7 &  7391.5 & 1.57 & 0.92 &       &       & 16.92 & 0.038 &   $<$1   \\
001224.02$-$102226.2 & 0.22822 &  2.69 &   52.3 &  3510.8 &  147.0 & 11071.3 & 1.81 & 1.10 & 17.79 & 17.83 & 16.61 & 0.037 &   3.73   \\
001247.93$-$084700.4 & 0.22006 &  1.95 &    0.0 &  2575.2 &   61.8 &  8422.0 & 1.70 & 1.39 &       &       & 16.92 & 0.032 &   $<$1   \\
002840.69$-$102145.0 & 0.32173 &  1.53 &    4.2 &   493.6 &   27.0 &  1608.8 & 1.58 & 1.35 &       &       & 18.70 & 0.037 &   $<$1   \\
004319.74$+$005115.3 & 0.30807 &  3.08 &   37.7 &  2149.9 &  176.6 &  7098.0 & 1.72 & 1.72 & 18.21 & 18.08 & 17.88 & 0.020 &   1.60   \\
\enddata
\tablecomments{\normalsize Col. 1, object name in J2000.0.
Col. 2, redshift given by the SDSS spectroscopic pipeline.
Col. 3, monochromatic luminosity $\lambda L_{\lambda}$ at 5100 \AA, In units of ergs~s$^{-1}$.
Col. 4, H$\beta$ narrow component flux; its typical error is 6\%.
Col. 5, H$\beta$ broad component flux; its typical error is 8\%.
Col. 6, H$\alpha$ narrow component flux; its typical error is 5\%.
Col. 7, H$\alpha$ broad component flux; its typical error is 5\%.
Col. 8--9, continuum slopes blueward and redward of 5600\AA, respectively
($f_{\lambda} = \lambda ^{-\indw}$); their typical error is 8\%.
Col. 10--11, GALEX calibrated magnitudes (AB) in the FUV ($f$) and NUV ($n$) bands,
respectively, uncorrected for Galactic extinction;
a ``$-$999'' is given for sources that are covered yet not detected by GALEX,
and a blank is given for 185 sources that are not covered by GALEX.
Col. 12, the SDSS \textit{i}-band magnitude (AB), uncorrected for Galactic extinction.
Col. 13, the Galactic color excess derived from Schlegel et al. (1998).
Col. 14, the integrated flux density at 20cm detected by FIRST, in unit of mJy;
the detection limit of 1 mJy is adopted as the upper limit for 336 objects
that are covered yet not detected by FIRST,
and a blank is given for 42 objects that are not covered by FIRST.
\emph{\textbf{Table 1 is now available in its entirety at
http://staff.ustc.edu.cn/\~{}xbdong/Data\_Release/blueAGN\_DR4/;
it will be available via the link to the machine-readable table on the MNRAS website.}}
}
\tablenotetext{a}{\normalsize ~Line flux in units of
$10^{-17}$~erg~s$^{-1}$~cm$^{-2}$.}
\end{deluxetable}

\begin{deluxetable}{lrrrrrrrrrr}
\tablecolumns{11}
\tabletypesize{\scriptsize}
\tablewidth{0pt}
\tablecaption{\ha\ and \hb\ Broad-Line Profile Measurements}
\tablenum{2}
\tablewidth{0pt}
\topmargin 0.0cm
\evensidemargin = 0mm
\oddsidemargin = 0mm

\tablehead
{
\colhead{} &
\multicolumn{5}{c}{broad \hb} &
\multicolumn{5}{c}{broad \ha}   \\
\colhead{} &
\multicolumn{5}{c}{-------------------------------------------------} &
\multicolumn{5}{c}{-------------------------------------------------}   \\

\colhead{Object} &
\colhead{FWHM} & \colhead{$\sigma_{\rm line}$} &
\colhead{$AI$} & \colhead{$SI$} & \colhead{$KI1$} &
\colhead{FWHM} & \colhead{$\sigma_{\rm line}$} &
\colhead{$AI$} & \colhead{$SI$} & \colhead{$KI1$}    \\
\colhead{(1)} & \colhead{(2)} & \colhead{(3)} & \colhead{(4)} & \colhead{(5)} &
\colhead{(6)} & \colhead{(7)} & \colhead{(8)} & \colhead{(9)} & \colhead{(10)} &
\colhead{(11)}
}
\startdata
000710.01$+$005329.0 &  9165.5 & 4798.8 & $-$0.16 & 0.93 & 0.38 &  9165.5 & 4798.8 & $-$0.16 & 0.93 & 0.38      \\
000834.72$+$003156.1 &  1953.0 & 1415.8 & $ $0.07 & 1.08 & 0.40 &  1953.0 & 1415.8 & $ $0.07 & 1.08 & 0.40      \\
000943.14$-$090839.1 &  5288.3 & 3406.5 & $-$0.04 & 1.05 & 0.47 &  4225.2 & 2236.0 & $-$0.01 & 1.05 & 0.53      \\
001224.02$-$102226.2 &  4294.2 & 4175.2 & $-$0.64 & 1.40 & 0.18 &  4710.4 & 3646.0 & $-$0.35 & 1.15 & 0.26      \\
001247.93$-$084700.4 &  3007.4 & 2488.0 & $-$0.10 & 1.13 & 0.35 &  3007.7 & 2488.0 & $-$0.10 & 1.13 & 0.35      \\
002840.69$-$102145.0 &  1371.8 & 1525.8 & $-$0.05 & 1.13 & 0.36 &  1371.8 & 1525.8 & $-$0.05 & 1.13 & 0.36      \\
004319.74$+$005115.3 & 12630.6 & 5827.4 & $-$0.02 & 1.01 & 0.71 & 12630.6 & 5827.4 & $-$0.02 & 1.01 & 0.71      \\
\enddata
\tablecomments{\normalsize Col. 1, object name in J2000.0.
FWHM and line dispersion ($\sigma_{\rm line}$) in unit of km~s$^{-1}$.
Asymmetry index ($AI$) defined as in De Robertis (1985);
shape index ($SI$) as in Boroson \& Green (1992);
kurtosis index ($KI1$) in Marziani et al. (1996); see \S3.2.3.
All parameters are derived from the model broad lines.
\emph{\textbf{Table 2 is now available in its entirety at
http://staff.ustc.edu.cn/\~{}xbdong/Data\_Release/blueAGN\_DR4/;
it will be available via the link to the machine-readable table on the MNRAS website.
}}
}
\end{deluxetable}

\clearpage

\begin{deluxetable}{lll}
\tablenum{3}
\tablewidth{0pt}
\tabletypesize{\small}
\tablecaption{Summary of Spearman Rank Correlation Tests \tablenotemark{a}}
\tablehead{
\colhead{} & \colhead{H$\alpha^{b}$/H$\beta^{b}$} & \colhead{$\alpha_{\lambda, 1}$} }
\startdata
$L_{5100}$              & $ $0.115 ($ 0.015           $) &   $ $0.457 ($1.0\times 10^{-6}$) \\
\lratio\ \tablenotemark{b}                & $-$0.051 ($ 0.281           $) &   $ $0.073 ($0.123            $) \\
\mbh\    \tablenotemark{b}                & $ $0.093 ($ 0.049           $) &   $ $0.187 ($6.8\times 10^{-5}$) \\
$R_{i}$  \tablenotemark{c}                 & $ $0.205 ($<10^{-4}         $) &   $ $0.044 ($0.381            $) \\
$P_{\mathrm 20cm}$ \tablenotemark{c}      & $ $0.209 ($<10^{-4}         $) &   $ $0.144 ($0.004            $) \\
FWHM \tablenotemark{d}                    & $ $0.078 ($ 0.098           $) &   $ $0.075 ($0.116            $) \\
$\sigma_{\rm line}$ \tablenotemark{d} & $ $0.099 ($ 0.037           $) &   $ $0.121 ($0.011            $) \\
skewness \tablenotemark{d}                & $-$0.203 ($1.5\times 10^{-5}$) &   $-$0.077 ($0.103            $) \\
kurtosis \tablenotemark{d}                & $-$0.047 ($ 0.324           $) &   $-$0.055 ($0.242            $) \\
$AI$     \tablenotemark{d}                & $-$0.020 ($ 0.673           $) &   $ $0.024 ($0.619            $) \\
$SI$     \tablenotemark{d}                & $-$0.002 ($ 0.974           $) &   $ $0.129 ($0.006            $) \\
$KI1$    \tablenotemark{d}                & $-$0.151 ($ 0.001           $) &   $-$0.070 ($0.138            $) \\
$KI2$    \tablenotemark{d}                & $ $0.047 ($ 0.320           $) &   $-$0.032 ($0.502            $) \\
\aox\    \tablenotemark{e}               & $-$0.089 ($ 0.147           $) &   $ $0.138 ($0.024            $) \\
$\alpha_{\lambda,1}$    & $-$0.109 ($ 0.022           $) &

\enddata
\tablenotetext{a}{\normalsize ~For each entry, we list the Spearman rank correlation
statistic ($\rho$) and the probability of the null hypothesis ($P_{null}$)
in parenthesis. If no censored data (upper limits) are present,
$\rho$ is equal to the Spearman's rank correlation coefficient $r_s$. }
\tablenotetext{b}{\normalsize ~The black hole masses are calculated using
the formalism presented in Vestergaard \& Peterson (2006, their equation 5)
with \bhb\ FWHM listed in Table 2;
Eddington ratios (\lratio) are calculated assuming
that the bolometric luminosity $L_{bol} \approx 9\lfive$
used for normal QSOs (Kaspi et al. 2000, Elvis et al. 1994).}
\tablenotetext{c}{\normalsize ~Using 404 sources covered by FIRST, of which 336 are upper limits.}
\tablenotetext{d}{\normalsize ~Computed based on the model H$\beta$ broad-line profiles.}
\tablenotetext{e}{\normalsize ~Using 268 sources matched with ROSAT;
\aox\ values are brought from Anderson et al. (2007).}
\end{deluxetable}


\end{document}